\documentclass[twocolumn,showpacs,preprintnumbers,amsmath,aps,amssymb,nofootinbib]{revtex4-1}
\usepackage{graphicx}
\usepackage{dcolumn}
\usepackage{amsfonts}
\usepackage{bm}
\usepackage{color}
\begin{document}

\title{Black Hole Thermodynamics from a Noncommutative Area Operator}

\author{S. P\'erez-Pay\'an}
\email{sinuhe@fisica.ugto.mx}
\author{M. Sabido}
\email{msabido@fisica.ugto.mx}

\affiliation{ Departamento  de F\'{\i}sica de la Universidad de Guanajuato,\\
 A.P. E-143, C.P. 37150, Le\'on, Guanajuato, M\'exico\\
 }%
 
\begin{abstract}
One key element to calculate thermodynamical properties for a black hole is the partition function. In this paper we have incorporated the idea of a two dimensional area in a noncommutative space and were able to calculate the partition function with such a spectra. Employing the canonical quantum statistics formalism we compute the temperature, entropy and time of evaporation for a Schwarzschild black hole.
\end{abstract}
\maketitle

\section{Introduction}

The proposal of reconciling quantum field theory and general relativity has been extensively studied, the aim of this endeavor is to develop a consistent quantum theory of gravity, despite all the effort a satisfactory theory  has not  been constructed. If one tries to quantize  general relativity with the tools of quantum field theory, it turns out that the theory has an ill ultraviolet behavior. Then constructing a quantum theory of gravity applying the well know technics of quantum field theory, seems to fail. This might imply that the geometrical properties of space-time changes at the Planck scale.  There have been several approaches to address this issue (being the most successful String Theory \cite{horo} and Loop Quantum Gravity \cite{ash}), an approach that has attracted a lot of attention is noncommutativity. 

The idea of a noncommutative space-time is not new, it had been studied from the  physics perspective by Snyder  \cite{snyder} and formal mathematical perspective was given by Connes \cite{connes}. During the last decade a lot of work has been done in noncommutative physics (in particular in connection to field theory, gravity \cite{ncsdg_a, mike}, cosmology \cite{nc_cosmo}, and as unification principle \cite{connes2}, etc).  One motivation for a noncommutative space-time, is based on an analogy with quantum mechanics. In quantum mechanics the phase space variables $x_i$ and $P_{x_i}$ are promoted to hermitic operators which obey the Heinsenberg algebra $[\hat{x}_i,\hat{P}_{x_j}]=i\hbar\delta_{ij}$, as it is known, the commutation relation implies the uncertainty principle $\Delta x\Delta P\ge\hbar/2$.  With this in mind one  defines a noncommutative space by promoting the  coordinates $x_i$ to hermitian operators $\hat{x}_{i}$ which obey the commutation relations
\begin{equation}
[\hat{x}^i,\hat{x}^j]=i\Theta^{ij},
\label{nc}
\end{equation}
where $\Theta^{ij}$ is an antisymmetric matrix with real and constant parameters with dimensions of $(\rm length)^{2}$.
The motivation for Snyder \cite{snyder} to introduce noncommutativity,  was that if one found a coherent description of space-time in which  the notion of a point is no longer valid at Planck scales, then the ultraviolet divergences of quantum field theory might disappear. Another argument can be made in favor of  noncommutativity; it is believed that  in quantum gravity  the measurement of length is limited to distances greater then the Planck length, Eq.(\ref{nc}) gives an uncertainty principle between the coordinates that can be interpreted as that limit.

On the other, hand if we think that black holes play a similar role in quantum gravity as the atom played in the development of quantum mechanics and quantum field theory, we may use black holes as a starting point for testing different constructions that include quantum aspects of gravity. Quantization of black holes was proposed in the pioneering work of Bekenstein \cite{bekenstein}, he suggested that the surface gravity its proportional to its temperature and that the area of its event horizon is proportional to its entropy. In his remarkable work he conjectured that the horizon area of non-extremal black holes plays the role of a classic adiabatic invariant. He concluded that the horizon area should have  a discrete spectrum with uniformly spaced eigenvalues
 \begin{equation}
A_n=\bar{\gamma}l_{p}^{2}n\label{ABeks},\qquad n=1,2,...
\end{equation}
where $\bar{\gamma}$ is a dimensionless constant. Also in \cite{ahluwalia} the author concludes that an important physical consequence of the conjectured universal equal area spacing \cite{hod}, is that the quantum gravity description of Schwarzschild black hole space-time is characterized by a quantum area operator which is equally spaced. 

In this paper we apply some of this ideas to obtain thermodynamical properties of a black hole. Using the canonical quantum statistics formalism on the area spectrum of a two dimensional noncommutative space, we are able to obtain themodynamical properties for a Schwarzschild black hole. Then we employ the black body power law to get the evaporation time for such system. As we will show the temperature, entropy and time of evaporation are dependent on the noncommutative parameter.

\section{The Noncommutative Area Operator}

There have been several approaches to study black holes \cite{ncbh}. For our purposes we need the spectrum of the black hole,  the spectrum can be calculated  using ideas from noncommutative
 quantum theory in a two dimensional Euclidean space \cite{vergara}. 

Beginning with the commutation relation
\begin{equation}
\left[\hat{x}_{1},\hat{x}_2\right]=i\hbar\Theta, \qquad \Theta>0,
\end{equation}
one can define the operators $\hat{a}=1/\sqrt{2\hbar\Theta}\left(\hat{x}_{1}+\hat{x}_2\right)$ and $ 
\hat{a}^{\dagger}=1/2\hbar\Theta\left(\hat{x}_{1}-\hat{x}_2\right)$,
which satisfy the commutation relations $\left[\hat{a},\hat{a}^{\dagger}\right]=1$.
With $\hat{a}$ and $\hat{a}^{\dagger}$ we construct the number operator, $\widehat{N}=\hat{a}^{\dagger}\hat{a}$,  that satisfies the algebra $[\widehat{N},\hat{a}^{\dagger}]=\hat{a}^{\dagger}$, and $[\widehat{N},\hat{a}]=-\hat{a}$.
In this sense, if $\widehat{N}\vert{n}\rangle=n\vert{n}\rangle$, we identify $\hat{a}^{\dagger}$ and $\hat{a}$ as creation and annihilation operators. In terms of the coordinates operators the number operator takes the form
\begin{equation}
\widehat{N}=\frac{1}{2\hbar\Theta}\left(\hat{x}_{1}^{2}+\hat{x}_{2}^{2}-\hbar\Theta\right),\label{NOp}
\end{equation}
defining the operator $\hat{A}/\pi\equiv\hat{x}_{1}^{2}+\hat{x}_{2}^{2}$, it follows from (\ref{NOp}) that
\begin{equation}
\frac{\hat{A}}{\pi}=2\hbar\Theta\left(\widehat{N}+1/2\right).
\end{equation}
Given that $\hat{A}/\pi$ is positive by definition, the operator $\widehat{N}$ has a minimal eigenvalue. As a consequence, $\hat{A}/\pi$ is quantized with levels equally spaced by the interval $2\hbar\Theta$,
\begin{equation}
\frac{\hat{A}}{\pi}\vert n\rangle=2\hbar\Theta(n+1/2),\quad n=0,1,...
\end{equation}
with $\vert n\rangle=((\hat{a}^{\dagger})^{n}/\sqrt{n!})\vert 0\rangle$.
One cannot consider $\sqrt{\hat{A}/\pi}$ as a distance operator, because in noncommutativity the notion of a point does not exist. However, the area operator can be related to the area of a circumference $A=\pi\left[(x_1)^2+(x_2)^2\right]$ where
his quantum counterpart is given by
\begin{equation}
\hat{A}=\pi\left[(\hat{x}_1)^2+(\hat{x}_2)^2\right].
\end{equation}
Thus the area has discrete spectra
\begin{equation}
A_n=2\pi\hbar\Theta(n+1/2).
\label{espectro}
\end{equation}
It is interesting that making the simple hypothesis of noncommutativity in two dimensions one obtains an equally space area spectrum that looks like the one predicted for black holes \cite{bekenstein}. The idea that black holes have an area operator with an equal space sprectrum was explored in \cite{ahluwalia} under a different context.

\section{black hole thermodynamics from the area operator}

Several authors, beginning with Bekenstein \cite{bekenstein}, have suggested that the energy levels of an isolated Schwarzschild black hole are $E_n=\sigma\sqrt{n}E_p, n\in\mathbb{N}$ and $E_p$ is the Planck Energy. Furthermore, in \cite{ahluwalia}, the author shows that the interpretation of the universal equal area spacing $\lambda_{P}^24\ln{(3)}$ (which is the unique value compatible with the area-entropy thermodynamic relations, with statistical arguments and the Bohr's correspondence principle \cite{hod}) implies a quantum harmonic oscillator like structure of space-time. These lines of reasoning lead to an area operator $\mathcal{A}=2\pi l_s+\frac{2[\ln(3)]^2}{\pi}l_{P}^{2}$ and a fundamental operator relation $\left[l_P,l_S\right]=i\lambda_{P}^2$, which reflects the quantum nature of black holes space-times.

Given that for a black hole $E\sim M$ and $A\sim M^2$, where $E$ is the energy, $M$ the mass and $A$ the area of the event horizon, we can  use a canonical  ensemble \cite{kastrup} and the spectra of the area operator Eq.(\ref{espectro}), to analyze the consequences that would have on a Schwarzschild black hole. In order to calculate the thermodynamical properties, we need the partition function. In our case $Z(x)=\sum_{n=0}^{\infty}e^{-\sqrt{n}x}$, promoting the sum to an integral and using the identity $\int^{\infty}_{0}(-ax^2-b/x^2)dx=\frac{1}{2}\sqrt{\pi/a}e^{-2\sqrt{ab}}$, we obtain the partition function
\begin{equation}
Z=\frac{\sqrt\pi x}{2}e^{-x^2/4},
\end{equation}
where $x=\sigma E_{p}\alpha(\Theta)\beta$, $\alpha^2(\Theta)=(\hbar\Theta)/8$ and $\beta$ the Boltzmann factor.

We know that the internal energy for a system is given by $\bar{E}=-\frac{\partial}{\partial{\beta}} \ln Z$,
solving for $\beta$ in terms of the Hawking temperature $\beta_H=8\pi Mc^2/E_{p}^{2}$, taking the positive root for the  $Mc^2 \gg E_p$ and setting $\sigma=1/4\pi$,
we can obtain a temperature that has a $1/\alpha^2(\Theta)$ dependence. Defining\footnote{All quantities with the label ${}^{(nc)}$, are quantities related to the noncommutative area operator.} $\beta^{(nc)}_{H}=\beta_H/\alpha^2(\Theta)$ we get,
\begin{equation}
\beta^{(nc)}=\beta^{(nc)}_{H}\left[1-\frac{1}{\beta^{(nc)}_{H}}\frac{1}{Mc^2}\right]\label{beta}.
\end{equation}
In order to calculate the entropy we use $S/{k}=\ln Z+\beta\bar{E}$, which in terms of the Hawking-Bekenstein entropy $S/k=4\pi(Mc^2/E_{p}^2)$ we arrive to

\begin{equation}
\frac{S^{(nc)}}{k}= \frac{S^{(nc)}_{HB}}{k}+\frac{1}{2}\ln\left[\frac{S^{(nc)}_{HB} }{k}\right]+\ln(\pi)-1+\mathcal{O}(S^{(nc)}_{HB})^{-1},
\label{entropia}
\end{equation}
where we have defined $S^{(nc)}_{HB}=S_{HB}/\alpha^2(\Theta)$
as the Hawking-Bekenstein entropy divided by the square of the noncommutative factor.

Another property that can be calculated is the evaporation time for the black hole. We know that the  evaporation time for a black hole is proportional to $M^3$, this can be seen from the black body power law $P=\sigma AT^4$, where $\sigma$, $A$ and $T$ are the Stefan-Boltzmann, area and temperature of the black hole respectively. In our model we can can identify the area from  (\ref{beta}) and the temperature from (\ref{entropia}). First,  $\beta$ is equal to the Hawking temperature divided by the noncommutative factor $\alpha^2(\Theta)$, so $T\sim\alpha^2(\Theta)/M$. From the relation for the entropy, we defined the entropy for our model as $S^{(nc)}_{HB}=S_{BH}/\alpha^2(\Theta)$, but we know that the Hawking-Bekenstein entropy is equal to $A/4\pi$, so our ``noncommutative area" will be $ A/4\pi\alpha^2(\Theta)\sim\ A/\alpha^2(\Theta)$. It turns out that we can cast these two quantities in terms of the commutative ones weighted by the noncommutative factor $\alpha^2(\Theta)$. Putting all the ingredients into the black body power law we have that the time of evaporation for this black hole is given by
\begin{equation}
\tau^{(nc)}=\frac{c^2M_{i}}{3\kappa\sigma}\frac{1}{\alpha^{3}(\Theta)}=\frac{\tau}{\alpha^{3}(\Theta)},
\end{equation}
where $c$ is the speed of light, $M_i$ is the initial mass of the black hole,  $\kappa=\hbar^4c^8/256\pi^3G^2k^2$ and $\sigma$ is the Stefan-Boltzman constant. We can see that the noncommuative time of evaporation is proportional to the commutative one divided by the cube of the noncommutative parameter $\alpha(\Theta)$.

We showed that the inclusion of the two-dimensional noncommutative area operator to the partition function modifies the thermodynamical quantities of interest. For the temperature and entropy the dependence on the noncommutative parameter manifest as the square of $1/\alpha(\Theta)$, while for time of evaporation the dependence comes into play as $1/\alpha^3(\Theta)$.

\section{Conclusions}

To probe aspects of quantum gravity, in particular noncommutative effects, black holes seem to be natural candidates. One key element to calculate thermodynamical properties is the partition function, in this paper we have incorporated the idea of a two dimensional area in a noncommutative space and were able to calculate the partition function with such a spectra. Employing the canonical quantum statistics formalism we were able to compute the temperature, entropy and time of evaporation for a Schwarzschild black hole. We found that these three quantities can be written in terms of the commutative ones weighted by the noncommutative factor $\alpha(\Theta)$. The first one is written in terms of the Hawking temperature and the dependence from noncommutativity appears as the inverse of the noncommutative factor $\alpha^2(\Theta)$. For the entropy we have a logarithmic correction and is weighted by the inverse of noncommutative factor. Finally the evaporation time $\tau^{(nc)}$ is expressed in terms of the commutative one divided by $\alpha^3(\Theta)$.
The noncommutative expressions found in this paper for the temperature and entropy have a very similar form to those found in \cite{mukherji,loop1}.

\section*{Acknowledgments}
 This work was partially supported by UG DAIP grant  and  CONACYT grants 62253, 135023. S. P. P. is supported by CONACyT graduate grant.

\end{document}